\documentclass[12pt]{article}
\usepackage{amsfonts}
\usepackage{amssymb}
\usepackage{latexsym}
\usepackage{graphicx}
\usepackage[english]{babel}
\begin{document}
\input epsf

\def\p{\partial}
\def\h{{1\over 2}}
\def\be{\begin{equation}}
\def\bea{\begin{eqnarray}}
\def\ee{\end{equation}}
\def\eea{\end{eqnarray}}
\def\d{\partial}
\def\la{\lambda}
\def\eps{\epsilon}
\def\bb{\bigskip}
\def\mm{\medskip}
\newcommand{\dm}{\begin{displaymath}}
\newcommand{\edm}{\end{displaymath}}
\renewcommand{\b}{\tilde{B}}
\newcommand{\gm}{\Gamma}
\newcommand{\ac}[2]{\ensuremath{\{ #1, #2 \}}}
\renewcommand{\ell}{l}
\newcommand{\z}{\ell}
\newcommand{\newsection}[1]{\section{#1} \setcounter{equation}{0}}
\def\bb{$\bullet$}
\def\Qbar{{\bar Q}_1}
\def\QPbar{{\bar Q}_p}

\def\q{\quad}

\def\bn{B_\circ}

\let\a=\alpha \let\b=\beta \let\g=\gamma \let\d=\delta \let\e=\epsilon
\let\c=\chi \let\th=\theta  \let\k=\kappa
\let\l=\lambda \let\m=\mu \let\n=\nu \let\x=\xi \let\r=\rho
\let\s=\sigma \let\t=\tau
\let\vp=\varphi \let\vep=\varepsilon
\let\w=\omega      \let\G=\Gamma \let\D=\Delta \let\Th=\Theta
                     \let\P=\Pi \let\S=\Sigma

\def\h{{1\over 2}}
\def\t{\tilde}
\def\r{\rightarrow}
\def\nn{\nonumber\\}
\let\bm=\bibitem
\def\Kt{{\tilde K}}
\def\b{\bigskip}

\let\p=\partial

\begin{flushright}
\end{flushright}
\vspace{20mm}
\begin{center}
{\LARGE  Tunneling into fuzzball states}
\\
\vspace{18mm}
{\bf  Samir D. Mathur\footnote{mathur@mps.ohio-state.edu} }\\

\vspace{8mm}
Department of Physics,\\ The Ohio State University,\\ Columbus,
OH 43210, USA\\
\vspace{4mm}
\end{center}
\vspace{10mm}
\thispagestyle{empty}
\begin{abstract}

String theory suggests that black hole microstates are quantum, horizon sized `fuzzballs', rather than   smooth geometries with  horizon. Radiation from fuzzballs can carry  information and does not lead to information loss.  But if we let a shell of matter collapse then it creates a {\it horizon}, and it seems that subsequent radiation {\it will} lead to information loss. We argue that the resolution to this problem is that  the shell  can tunnel to the fuzzball configurations. The amplitude for tunneling  is small because we are relating two macroscopically different configurations, but the number of states that we can tunnel to, given through the Bekenstein entropy, is very large. These small and large numbers can cancel each other, making it possible for the shell to tunnel into fuzzball states before a significant amount of radiation has been emitted. This offers a way to resolve the information paradox.

\end{abstract}
\newpage
\setcounter{page}{1}

\section{The information paradox}

In recent years we have learnt some intriguing lessons about the nature of quantum gravity. In this venture a key role has been played by the mysteries posed by black holes. 

Black holes are puzzling objects. They have an enormous entropy $S_{bek}$, but where are the states that account for this entropy? The metric of the black hole seems to be unique (`black holes have no hair'), which would suggest an entropy  $S=\ln 1=0$. We have begun to understand the resolution of this puzzle: black holes are {\it not} as classical as we used to think. Quantum effects smear the information in the hole all over a horizon sized `fuzzball'. There are $e^{S_{bek}}$ such fuzzball states, and the traditional geometry of the hole can be understood only as some effective description of this ensemble of microstates \cite{twocharge,threecharge}.

\begin{figure}[h]
\begin{center}
\includegraphics[width=28pc]{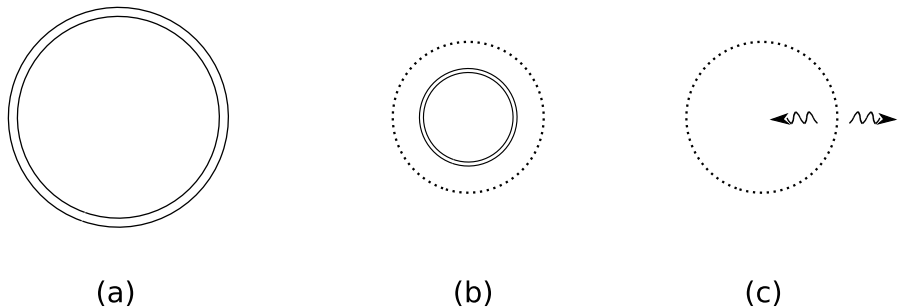}
\end{center}
\caption{\label{fone}Formation and evaporation of a black hole, in the traditional picture.}
\end{figure}

The second (and more serious) puzzle with black holes is the information paradox (fig.\ref{fone}).  Suppose a shell of matter collapses under its own gravity. It passes through its horizon radius, apparently described to good accuracy by classical dynamics. No large curvatures are involved at the horizon, so it would seem that quantum effects cannot impede the collapse of the shell through the horizon. But now particle-antiparticle pairs are created in the vacuum region around the horizon and the hole slowly radiates away its mass, in a way that violates quantum unitarity \cite{hawking}.

How do we escape this problem and save quantum theory? In this article we will argue that the large entropy characteristic of black holes can lead to a surprising quantum effect. In the everyday world around us the evolution of macroscopic objects obeys classical dynamics; even though one can `quantum tunnel' from one macroscopic configuration to another, the action for this is so small that we can comfortably ignore such processes.  The black hole is also macroscopic, and tunneling amplitudes between states will be  small, but the {\it number} of states that are available to tunnel to is $e^{S_{bek}}$, a very {\it large} number. We will observe that these small and large numbers compensate each other, and allow the collapsing shell to tunnel to a linear combination of fuzzball states. These fuzzball states do {\it not} have a horizon, and so radiate in a manner that does not lead to information loss. Thus the large entropy of the hole can be the key to resolving the information paradox.

\section{The structure of microstates}

\begin{figure}[h]
\begin{center}
\includegraphics[width=28pc]{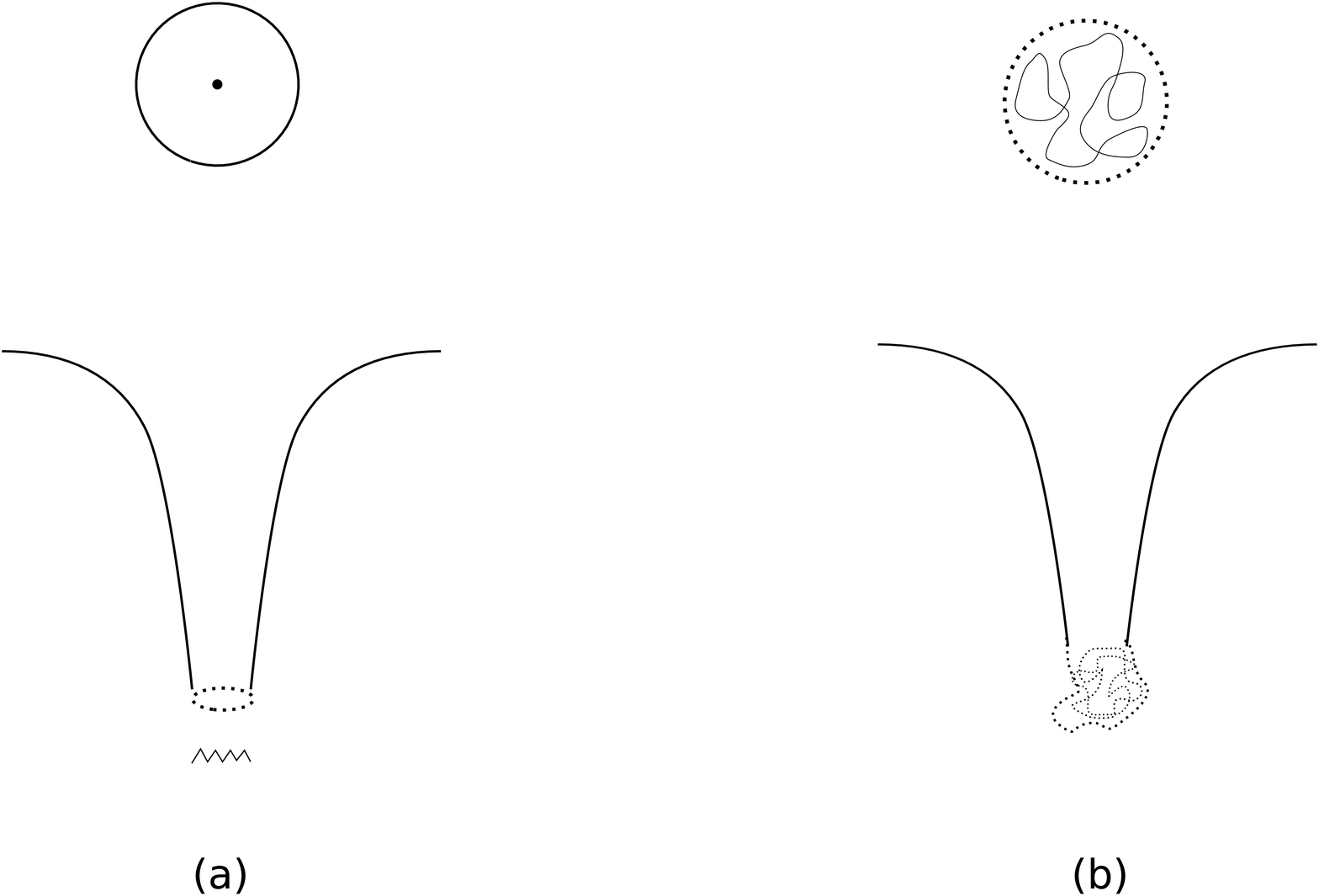}
\end{center}
\caption{\label{ftwo} (a) The traditional black hole has a throat, horizon and singularity. (b) The actual microstates are fuzzballs, for which the throat ends in a quantum fuzz.}
\end{figure}

The fuzzball states will play a crucial role in our argument, so let us first discuss their nature in more detail. In string theory, we can take the fundamental quanta of  the theory (strings, branes etc.) and put them together to try and make black holes.  It is known that the entropy of black holes can be reproduced by a count of microscopic string states \cite{count}. Our interest now is in the {\it structure} of these states.  Suppose we take $N$ quanta, with  total mass $M$ and  charge $Q$.  If $N$ is large we might think that the the system is classical, described by the traditional black hole metric with mass $M$ and charge $Q$ (fig.\ref{ftwo}(a)). On the other hand we can look at the detailed construction of the string theory  states made by the $N$ quanta. It turns out that {\it none} of the states made by these quanta look like the traditional black hole. For one thing, the quanta carry a `spin', and choosing an orientation for these spins breaks spherical symmetry. More importantly,  the  size of the bound state of these $N$ quanta is not of order planck length $l_p$. Instead it grows as $N^\alpha l_p$, with $\alpha$ such that we get  horizon sized `fuzzball' states (fig.\ref{ftwo}(b)) \cite{twocharge,threecharge}.

The property of the fuzzball states that is crucial for us is that {\it there is no horizon with a vacuum in its vicinity}. Recall that information loss occurs when particle pairs are pulled out of the vacuum by the gravitational field at a horizon. By contrast, radiation from the fuzzball carries out information just like the radiation from a piece of burning coal. 
 The problem that we are facing is that we have started with a collapsing shell, and this does {\it not} look like a fuzzball state. As the shell shrinks  it can creates a horizon which has a {\it vacuum} in its vicinity. In that case we would  still face the information paradox, unless we can find some way of changing the collapsing shell to a fuzzball state.

\section{Tunneling}

We make two preliminary observations that will be important:

\bigskip

(i) The  fuzzball solutions are time {\it independent} states of the hole. By contrast, the collapsing shell is a very special, very low entropy, time {\it dependent} configuration. In a general statistical system, such a special initial state will evolve towards a generic state, and only then will the system be characterized by its full entropy.

\bigskip

(ii)  There are {\it two} timescales associated to the black hole. One is $t_{cross}\sim GM$, the `crossing time' for light across distances of order the horizon scale. The infall of the shell occurs on this timescale. The other is the much longer Hawking evaporation timescale, $t_{evap}\sim ({M\over M_{pl}})^2t_{cross}$. It is known that the entropy of all the Hawking radiation from the hole is about 30\% larger than the Bekenstein entropy of the hole \cite{extra}. This fact has the following interesting implication. We can allow the shell to pass through its horizon, and even emit some quanta from this horizon containing {\it no} information, as long as the shell reaches a generic fuzzball configuration before $\sim $30\% of the radiation has been emitted. (Subsequent radiation from the fuzzball will of course carry information.) 

\bigskip

What we need to understand now is how the shell can transform into a fuzzball. We conjecture that this can take place through a process of {\it tunneling} between the shell state and the fuzzball configuration.The action for tunneling can be estimated to be of order
\be
S_{tunnel}\sim {1\over G}\int \sqrt{-g} R =\alpha GM^2, ~~~\alpha=O(1)
\label{six}
\ee
where we have used the black hole length scale $GM$ to estimate the curvature scale in $S_{tunnel}$. Thus the amplitude for tunneling is
\be
{\cal A}\sim e^{-S_{tunnel}}\sim e^{-\alpha GM^2},
\label{four}
\ee
a  very {\it small} quantity. But the {\it number} of fuzzball states that we can tunnel to is given by 
\be
{\cal N}\sim e^{S_{bek}}\sim e^{GM^2}, 
\label{three}
\ee
a  very {\it large} number. 

To see how these small and large numbers can play off against each other, consider a toy model. A  particle  is trapped in a potential well in a state  $|\psi_0\rangle$,  but is allowed to tunnel to  neighboring wells  with a very small amplitude  $\sim e^{-S_{tunnel}}$. Let the space in which the particle lives have a very large dimension $d$ (to reflect the large dimension of the space of fuzzball solutions); thus the number of  neighboring wells will be ${\cal N}=2^d\gg 1$. Writing $|\psi_i\rangle$ for the state in well $i$, the state of the particle at time $t$ can be written as
\be
|\psi(t)\rangle=C_0(t)|\psi_0\rangle+\sum_{i=1}^{\cal N} C_i(t)|\psi_i\rangle
\ee
We start at time $t=0$  with  $C_0=1, C_i=0$. 
Then
\be
{dC_i\over dt}\sim e^{-S_{tunnel}}, ~~~C_i\sim e^{-S_{tunnel}} t\sim e^{-S_{tunnel}}
\ee
(As is usual in tunneling problems, we have dropped polynomial prefactors and kept only the dominating exponential.) Thus
\be
{d \over dt} \sum_{i=1}^{\cal N} |C_i|^2\sim {\cal N}e^{-2S_{tunnel}}
\ee
Since $|C_0|^2+\sum_i |C_i|^2=1$, the probability for the particle to be in the initial well becomes small after a time
\be
\Delta t\sim  {\cal N}^{-1} {e^{2S_{tunnel}}}
\ee

Returning to the black hole we note that  since 
$\ln {\cal N}=S_{bek}$ and  $S_{tunnel}$ are of the same order, it is possible that ${\cal N}^{-1} {e^{2S_{tunnel}}}$ is {\it not} exponentially large. 
Thus it may be possible for the the collapsing shell to tunnel to a superposition of fuzzball states in a time 
\be
t_{stabilization}\lesssim t_{evap}
\label{one}
\ee
after which information  will start being radiated out from the fuzzball state.

\bigskip

Finally, we discuss the technical computation of $S_{tunnel}$ for fuzzball solutions. All fuzzball solutions have the same  mass $M$ and charge $Q$.   But the details of the solutions are characterized by additional {\it dipole } charges $D_1, D_2, D_3, \dots$ which total to zero, but whose locations determine the structure of the fuzzball state \cite{dipole}. Nonextremal states have orbifold singularities and conical defects linking the insertions of these $D_i$. The $D_i$ are held apart at fixed distances by fluxes of electric and magnetic fields. To move between fuzzball states, or between a shell state and a fuzzball state,  imagine a process where we have some dipoles $D_i$, and we tunnel to a configuration with an additional pair of dipoles $D, \bar D$. This process is like pair creation of charges in an external electric field, and its action can be computed. The action is $\sim {GM^2}$ if the separation of the $D, \bar D$ is $\sim GM$, and less if they are closer. Thus we expect $S_{tunnel}\lesssim GM^2$, which implies (\ref{one}).

\bigskip

In summary, we have found an intriguing quantum behavior for a shell that falls through its horizon radius. String theory provides us with $e^{S_{bek}}$ nonsingular `fuzzball' configurations that represent the microstates of a black hole. Once the shell reaches  a size $\lesssim GM$, the action  to tunnel to these other configurations is 
$\sim GM^2$, still a large number.  But the smallness of the tunneling amplitude is compensated by the large entropy $S_{bek}= GM^2$ of states that the shell can tunnel to. Interestingly,  this observation uses the curious fact that the black hole entropy is given by the {\it geometrical size} $\sim GM$ of the hole. The large entropy which distinguishes black holes from other objects has created an unusual quantum effect, and suggests a possible resolution of the information paradox. 

\section*{Acknowledgements}

 I would  like to thank Steven Avery, Borundev Chowdhury and Jeremy Michelson for helpful comments.  This work was supported in part by DOE grant DE-FG02-91ER-40690.

\end{document}